# Measuring Mars Atmospheric Winds from Orbit
A White Paper submitted 15 July 2020 to the
Planetary Science and Astrobiology Decadal Survey 2023-2032

**Authors:**
Scott Guzewich (NASA GSFC, scott.d.guzewich@nasa.gov, 301-286-1542)
J.B. Abshire (NASA GSFC)
M.M. Baker (Smithsonian Institution)
J.M. Battalio (Yale University)
T. Bertrand (NASA ARC)
A.J. Brown (Plancius Research)
A. Colaprete (NASA ARC)
A.M. Cook (NASA ARC)
D.R. Cremons (NASA GSFC)
M.M. Crismani (NASA GSFC)
A.I. Dave (NASA ARC)
M. Day (UCLA)
M.-C. Desjean (CNES)
M. Elrod (NASA GSFC)
L. K. Fenton (SETI Institute)
J. Fisher (U. Wollongong, Australia)
L.L. Gordley (GATS, Inc.)
P. O. Hayne (U. Colorado Boulder)
N.G. Heavens (Space Science Institute)
J.L. Hollingsworth (NASA ARC)
D. Jha (MVJCE, India)
V. Jha (NASA ARC)
M.A. Kahre (NASA ARC)
A.SJ. Khayat (NASA GSFC)
A.M. Kling (NASA ARC)
S.R. Lewis (Open University, UK)
B.T. Marshall (GATS, Inc.)
G. Martínez (USRA/LPI)
L. Montabone (Space Science Institute)
M.A. Mischna (JPL/Caltech)
C.E. Newman (Aeolis Research)
A. Pankine (Space Science Institute)
H. Riris (NASA GSFC)
J. Shirley (JPL/Caltech)
M.D. Smith (NASA GSFC)
A. Spiga (LMD, France)
X. Sun (NASA GSFC)
L.K. Tamppari (JPL/Caltech)
R.M.B. Young (NSSTC/UAEU)
D. Viúdez-Moreiras (CSIC-INTA, Spain)
G.L. Villaneuva (NASA GSFC)
M.J. Wolff (Space Science Institute)
R. J. Wilson (NASA ARC)

**Signatories:**
D. Banfield (Cornell University)
A. Brecht (NASA ARC)
S. Byrne (University of Arizona)
M. Chojnacki (University of Arizona)
E. Mason (NASA GSFC)
I. B. Smith (York University and PSI)

**Key Point**
Wind is the process that connects Mars' climate system. Measurements of Mars atmospheric winds from orbit would dramatically advance our understanding of Mars and help prepare for human exploration of the Red Planet. Multiple instrument candidates are in development and will be ready for flight in the next decade. We urge the Decadal Survey to make these measurements a priority for 2023-2032.

# I. Introduction

> We, the authors and signatories of this white paper, advocate for global atmospheric wind observations from Mars orbit within the next decade of NASA planetary science. This white paper demonstrates the science and engineering case for measuring Mars atmospheric winds (i.e. vector-resolved, horizontal winds from the surface to ≥80 km altitude at ≤5 km vertical and ≤300 km horizontal resolution) from orbit.

Wind measurements remain a critical gap in our understanding of Mars' climate and atmospheric processes and have been highlighted by the new Mars Exploration Program Analysis Group (MEPAG) goals document as a high priority for Mars science [1]. Wind is one of the predominant forces that has shaped Mars' surface over the last 3 billion years, depositing, burying, and eroding an extensive sedimentary and climatic record comprising sand, dust, and ice. In addition, wind is the agent of transport for water vapor, dust, trace gases, and to a significant extent, heat, around the planet. Understanding wind and the atmospheric circulation is necessary to understand the present and past water, dust, and $CO_2$ cycles.

We highlight the relevance of these measurements to future human and robotic exploration of Mars. Understanding wind patterns helps close strategic knowledge gaps identified for human exploration of Mars, reduces risk in entry, descent, and landing of spacecraft and their launch from the surface, and helps elucidate and constrain the environmental impacts human exploration may have on the pristine martian environment [2].

We demonstrate that independent, but complementary, measurement techniques can be used to profile wind from the surface to ≥80 km from orbit and many of these techniques can operate during dust storm conditions, a period during which wind measurements are of particular interest for elucidating mechanisms of storm growth and decay. New and existing instruments, which apply these techniques, are in development at multiple US institutions and will be ready for flight on the next Mars orbiting science mission.

# II. Science Motivation

**Wind plays a dominant role in geologic and climatic processes on modern-day Mars**. The exchange of mass and momentum between the atmosphere and surface represents a dynamic system that is impacted by the four-dimensional wind field. Surface winds mobilize sediment, modify surface albedo and loft dust into the atmosphere, thereby impacting insolation and surface and atmospheric heating. As on Earth, the circulation - which is directly provided by wind measurements - is responsible for transport of tracers from dust to water to trace gases such as methane [3,4,5]. Despite the essential role wind plays in the evolution of Mars as a planetary system, very few direct measurements of wind have been made. Additional high-quality measurements are needed to address these knowledge gaps and advance our understanding of the planet.

*Atmospheric Circulation, Transport, and Dust Storms*

Winds regulate Mars' climate through the transport of heat, volatiles, and aerosols. Yet our existing knowledge of the circulation is largely based on numerical climate models whose wind fields remain poorly validated (e.g. 6, 7). Large-scale patterns that have been predicted by the models and expected from theory include the Hadley circulation, trade winds, monsoon jets, western boundary currents, thermal tides, and baroclinic storms, all of which vary on timescales from hours to seasons.

Realistic, time-evolving wind fields are essential to successful back-tracking and attribution to surface locations of any spatially-variable species transported in the martian atmosphere. **Wind is critical to understanding the sources, sinks, and movement of methane and other astrobiologically-interesting trace gases**. Water undergoes a seasonal cycle between the atmosphere, regolith, and ice caps that is dependent on wind, impacting both surface-atmospheric exchange and transport. Dust storms are the greatest cause of sol-to-sol and interannual variability in Martian weather and climate. Wind is crucial to the development of dust storms, in two main respects: dust storms are generated via lifting from the surface by wind stress; and lofted dust is then transported by winds above the surface, with significant quantities of dust reaching 80 km or higher in the case of global storms. Measuring the 4D wind field in the lower atmosphere would help reveal the positive feedbacks between dust lifting, radiative heating and wind stress that may be crucial for onset and early growth of major storms (e.g. 8, 9, 10, 11). Wind measurements are called out by [12].

*Geology and Aeolian Processes*

For at least the past 3 Ga, the surface and sedimentary rock records of Mars have been sculpted predominantly by wind-driven physical weathering [13], in some locations removing up to hundreds of meters of material (e.g., 14, 15, 16). These sedimentary accumulations are a way to directly study the early climate of the planet. Sediment-laden wind abrades the surface, exhuming aqueous, volcanic, and possibly biosignature-bearing strata, the desired targets of most landed spacecraft (e.g., 17, 18). However, little is known about where the surface is being actively abraded, in part because surface winds have only been measured by lander instruments at a few locations, and without the time-resolution and duration required to study surface abrasion rates. Measurements of dune and ripple migration give some indication of where sediment supply is high and transport active enough to cause abrasion [19, 20], but a basic characterization of the physics of aeolian sediment transport is still incompletely established. Rover observations of sediment motion on Mars [21] identified sand transport at wind speeds below the theoretical threshold of motion, demonstrating that equations derived from experiments on Earth insufficiently describe martian aeolian transport [22, 23, 24]. Before fully taking advantage of the sedimentary rock record on Mars as a tool for studying the history of the planet, we need to understand the fundamental physics of aeolian transport. Coupled wind measurements and surface observation are required.

*MEPAG Goals and Mission Studies*

Wind measurements from orbit would provide critical input for addressing three of four MEPAG Science Goals [1], which highlights global wind measurements as high priority science to address Goal II Investigations regarding the current martian climate. Wind observations contribute to Goal III Investigations regarding the geologic expression of surface-atmosphere volatile exchange and Goal III Investigations relating to aeolian processes and dust generation and lifting mechanisms. Goal IV, relating to preparing for human exploration, lists wind observations as being integral to human exploration, specifically the high priority need for global wind observations from orbit.

Science Analysis Groups (SAGs) chartered by MEPAG since the last Planetary Science Decadal Survey have necessitated inclusion of wind observations for future missions. The Next Mars Orbiter SAG (NEX-SAG) [25] stated,

> *"Observation of wind velocity is the single most valuable new measurement that can be made to advance knowledge of atmospheric dynamic processes."*

The Ice and Climate Evolution SAG (ICE-SAG) include wind measurements as one of the highest priorities for understanding the transport of volatiles and dust between atmospheric and surface reservoirs [26]. Both SAGs and [1] emphasize that particular value is given to simultaneous measurements of wind, temperature, dust and water ice opacities, and other atmospheric state variables so that atmospheric transport, fluxes, and sources and sinks can be identified.

*Vision and Voyages in Planetary Science, 2013-2022*

The most recent Planetary Decadal Survey [27] recognized the importance of wind measurements on Mars and their significance to the top-level objectives of NASA planetary science research and planetary bodies other than Mars. In discussing outstanding questions in Mars climate research in particular, the report highlighted, "What is the four-dimensional wind structure of the Martian atmosphere from the surface boundary layer to the upper atmosphere?"

*Existing Wind Measurements*

<u>Landers</u>: Wind sensors have been included in the payload of every mission sent and scheduled to the surface of Mars, except for the Mars Exploration Rovers [28, 29, 30]. The wind sensors onboard the Viking Lander 1, Pathfinder, and Mars Science Laboratory (MSL) missions all experienced problems that prevented or limited reliable measurements. As a result, the longest record of wind measurements is still provided by the Viking Lander 2, with almost two complete Martian years of continuous measurements. However, wind speed and direction are strongly influenced by local and regional topography, thus are not always representative of the large-scale circulation. For example, recent results by InSight in Elysium Planitia show that the diurnal cycle of wind direction is influenced both by the global Hadley circulation and by flows induced by the nearby gentle regional slope (e.g., 31). Hence global atmospheric wind observations from Mars orbit are needed both to understand global transport patterns and to better interpret and place in context local-scale flows.

<u>MAVEN</u>: The Mars Atmosphere and Volatiles EvolutioN (MAVEN) Neutral Gas and Ion Mass Spectrometer (NGIMS) began measuring the winds of the upper atmosphere (~125-300 km) in 2016 and continues making wind measurements monthly by sweeping the instrument back and forth across the ram pointing [32]. In June 2020, MAVEN began raising its periapsis and the monthly wind measurements were no longer able to be made at the lower altitudes.

<u>Ground-based Telescopes</u>: Occasional ground-based single-dish and interferometric observations of the disk of Mars have provided line-of-sight wind velocities. High-resolution heterodyne millimeter spectroscopy observed the doppler-shifted spectral line cores of carbon monoxide (CO) (well-mixed in Mars' atmosphere), and provided wind velocities in the middle atmosphere where the spectral line cores are formed [33, 34, 35]. In addition, [36] conducted heterodyne infrared observations of the non-LTE $CO_2$ emission feature in the atmosphere and presented zonal wind measurements across the disk of the planet. However, these millimeter observations were limited by their poor spatial resolution (typically 10 arcsec/~3400 km [35]) and very low temporal coverage.

<u>Cloud tracking from the surface and orbit</u>: Cloud tracking retrieves atmospheric winds by imaging the positions of transient atmospheric phenomena (e.g., water or dust clouds) over time, and using their displacement between images to retrieve cloud motions, assuming that clouds are passive tracers of the flow which reflect the underlying winds. It has been used to measure winds at specific local times, altitudes, and latitudes, e.g. [37] using MGS-MOC, [38] using THEMIS-VIS, and [39] using MEx-VMC. However, cloud tracking is limited to daytime observations only and the sparse nature of martian clouds limits its utility to observe globally with sufficient vertical and spatial resolution.

*Why Atmospheric Modeling is Insufficient*

Information about the state of the atmosphere may be obtained from two sources: spacecraft measurements observe the state of the atmosphere at the time of the measurements, but are limited in both spatial and temporal coverage, and general circulation models (GCMs) which are a means of filling in the gaps within the observational record. Much work has gone into the development of these models and, while they serve as a useful tool for understanding atmospheric behavior in general, they suffer from a lack of observational validation. The variety of unique Mars GCMs are known to differ in their predictions (e.g., [6]), in particular those of wind directions and magnitudes which may be significantly inaccurate. Their differences and lack of grounding in observational data limit their ability to provide insight into Mars atmospheric and climate processes and provide useful instantaneous forecasts of weather – essential for future landed robotic and human activity. **The current state of numerical modeling is, by itself, insufficient to meet the needs of the Mars exploration program going forward.** Furthermore, the lack of validated models of Mars' present day circulation casts great doubt on the validity of such models' predictions for the circulation in Mars' recent or ancient past, which are heavily relied upon for interpreting Mars' past geology and habitability [40, 41, 42].

To mitigate these weaknesses in numerical modeling, data assimilation (DA) frameworks have been adapted from the terrestrial weather forecasting community, which blend the strengths of both direct observation and numerical modeling (e.g., 43, 44, 45, 46, 47, 48). Under a generalized DA framework, the numerical model 'background state' of the atmosphere is augmented with available observations to 'steer' an analysis of the true atmospheric state towards the values of the more accurate observations. At present, the atmospheric fields most regularly observed from orbit are temperature and dust opacity (from TES [49] and MCS [50, 51]). However, in the absence of networks of surface pressure measurements, which remain vital on Earth but unlikely on Mars, direct wind measurements are essential to constrain the absolute modeled wind magnitude. Assimilation of wind data, both magnitude and direction, is one key new data source that can markedly improve DA forecasts at Mars, as it provides a direct, rather than second-order, correction to the modeled wind field.

## III. Importance for Future Human and Robotic Spaceflight

NASA has identified a variety of "Strategic Knowledge Gaps" (SKGs) for human exploration of Mars. As described by [1] Goal IV and [2], closing SKGs related to climatological risks for human missions is a high-priority for Mars exploration. Atmospheric conditions such as the vertical temperature profile, the distribution of aerosols, and the four-dimensional wind field pose a risk to spacecraft EDL (e.g., aerodynamic maneuvering) and surface operations (e.g., damage to instruments from wind-blown material). As the level of acceptable risk is much lower for crewed missions than for robotic missions, the need to address outstanding knowledge gaps regarding the dynamics and variability of Mars' atmosphere becomes increasingly important. Winds significantly impact the simulated dynamics of parachutes and landers [52]. Furthermore, a crewed mission to Mars will include novel mission components such as in situ resource utilization (ISRU). Additional understanding of atmospheric conditions are needed to guide mission architecture and engineering design, as well as to mitigate risks to landed instruments and human explorers [53].

Previous robotic missions to Mars and human missions to the Moon have demonstrated the risks posed by atmospheric particulates to landed instruments and astronaut equipment. [1] states: "particulates can affect engineering performance and lifetime of hardware and infrastructure." Apollo spacesuits were coated and abraded by dust particulates, which threatened the integrity of suits and astronaut safety. On Mars, airborne fine-grained dust poses a threat to human explorers and their equipment. Accumulation and removal of wind-blown material also poses a unique challenge for solar power.

During crewed surface missions, ISRU systems may be used for crew consumables and ascent vehicle propellent. As such, there is a critical need to "understand the resilience of atmospheric In Situ Resource Utilization processing systems to variations in martian near surface environmental conditions" [1]. In this sense, additional data on the state and variability of the surface regime (e.g., wind speed, wind direction, and pressure), and resultant wind-driven sediment flux will play a critical role in guiding instrument design and placement [54]. Besides the direct hazards posed to instruments and humans, surface winds may also introduce unforeseen consequences for planetary protection, as wind-driven transport of Earth-based microbes could provide an opportunity for forward contamination on Mars.

## IV. What Wind Measurements are Needed

Our *science baseline* observation parameters and cadence are governed by the most stringent science questions that can be addressed by wind measurements from orbit, the requirements to validate atmospheric models, stated MEPAG objectives, and plausible technical feasibility within the next decade. It is as follows: vector-resolved horizontal winds from the surface to ≥80 km altitude at ≤5 km vertical and ≤300 km horizontal resolution with a precision of ≤5 m/s over the entire planet for 1 Mars year at ≥2 local times-of-day. Observations meeting the science baseline would fully address high priority science goals given by [1], substantially improve atmospheric models and their predictive capabilities, and reduce risk to future human exploration of Mars. As stated by MEPAG [1], we emphasize that these observations are most valuable when made in conjunction with other atmospheric state variables such as temperature, dust and water ice opacities, and trace gas (e.g., $H_2O$) mixing ratios.

There are plausible ways in which the science baseline may be exceeded in certain aspects during the next decade. For instance, higher vertical (e.g., ≤2 km) and/or horizontal (≤50 km) resolution in the lower atmosphere and boundary layer would identify medium- and fine-scale processes associated with topography and transient phenomena. Observations over multiple Mars years and during a global dust storm would be highly valuable to constrain variability and understand how Mars' great dust storms impact climate processes such as loss of water to space [55, 56, 57]. Observations at 4-8 local times-of-day are necessary for a system with a strong diurnal cycle, which is insufficiently characterized in existing Mars atmospheric observations [58].

## V. Available Instrumentation to Make These Measurements

### *LIDAR*

Lidar can measure martian winds by detecting the range-resolved Doppler shift of the laser backscatter profile of dust/aerosol along the laser beam line of sight. It can measure winds both day and night and during dust storms. It is most sensitive to the lowest ~40 km of atmosphere and insensitive above 60 km. The Mars lidar for global climate studies from orbit (MARLI) is a direct-detection Doppler wind lidar that measures the wind speed along the laser's line-of-sight in addition to dust and water ice extinction [59]. MARLI has been developed through PICASSO and MATISSE program funding to NASA GSFC and was planned to reach TRL 6 in June 2020, prior to the COVID-19 reduction in work. Wind speed is retrieved from the surface to ~40 km with measurement standard deviations under 5 m/s. This is even improved in dust storm conditions (standard deviations as low as 2 m/s). Through the cross-polarization of the laser backscatter, dust and water ice extinction can be distinguished from the surface to ~40 km with <10% relative error, allowing the continuation of the long record of atmospheric dust and water ice opacities over the last 20+ years. Critically, MARLI has a higher vertical resolution, 2 km, than most Mars atmospheric observations to-date (~5-10 km), which would address important science questions regarding the atmospheric circulation and vertical mixing of dust and water ice.

### *Sub-mm Sounder*

Submillimeter limb sounding can measure atmospheric winds by observing the line-of-sight position (Doppler shift) and shape (width, strength) of submillimeter thermal emission lines from molecules including $H_2O$, CO isotopologues and (potentially) other species [60]. It can measure day and night, is insensitive to dust storms, and is most sensitive from ~10 km to 100 km altitude. In the past several years, a sub-mm sounder has been optimized for use at Mars to obtain global atmospheric wind fields [61]. The instrument is TRL 6, except for SEU testing on the ASIC. One instrument with one articulating antenna or two antennas to view the same atmospheric mass from different angles can be used to retrieve wind vectors as a function of altitude. Wind profiles are achievable from ~10 km - 100+ km with a trade space in center frequency, vertical precision and resolution, scan rate, vertical scan distance, and averaging to achieve an optimized data acquisition approach. Sub-mm sounding is effective at measuring the winds in the middle and upper Mars atmosphere, which influence the lower atmosphere and are critical to understanding the overall circulation and the physics controlling it. A sub-mm sounder provides the information necessary to constrain atmospheric models and the physics within them.

*Doppler Wind and Temperature Sounder*

Gas filter correlation radiometry can be utilized to simultaneously measure the Doppler shift and linewidth of emission spectra toward the Mars limb, to infer both wind vectors and kinetic temperature [62]. This approach uses a cooled IR camera to observe the limb through an onboard gas filter cell, and provides observations of atmospheric dynamics from 5 km (or lower during periods of low dust storm activity) up to 150 km, depending on the chosen target gas(es). Winds can be measured with 5 m/s accuracy. The $\nu_3$ band of ozone has potential for Mars wind measurements from <10 km to over 60 km during day *and night observations*. Daytime measurements could be extended to altitudes of 100 km or more with the additional selection of mixed $CO/CO_2$ channels. SWaP estimates assume a dual channel ($O_3$ and $CO/CO_2$) system, providing limb observations from ~5 km up to 100+ km, with 5-km vertical resolution. Specific instrument elements, for example the ozone gas cell, are being built and tested to evaluate design performance, with a full single-channel prototype being developed for a terrestrial flight demonstration. These activities are aimed at raising the TRL of the instrument from TRL 4 to TRL 6.

## VI. Conclusion

Wind is the process that connects Mars' climate system. Measurements of Mars atmospheric winds from orbit would dramatically advance our understanding of Mars and help prepare for human exploration of the Red Planet. Multiple instrument candidates are in development and will be ready for flight in the next decade. We urge the Decadal Survey to make these measurements a priority for 2023-2032.

## VII. References Available at: https://bit.ly/2Bpf1bg

**References for "Measuring Mars Atmospheric Winds from Orbit"**

[1] MEPAG (2020), Mars Scientific Goals, Objectives, Investigations, and Priorities: 2020. D. Banfield, ed., 89 p. white paper posted March, 2020 by the Mars Exploration Program Analysis Group (MEPAG) at https://mepag.jpl.nasa.gov/reports.cfm.

[2] P-SAG (2012) Analysis of Strategic Knowledge Gaps Associated with Potential Human Missions to the Martian System: Final report of the Precursor Strategy Analysis Group (P-SAG), D.W. Beaty and M.H. Carr (co-chairs) + 25 co-authors, sponsored by MEPAG/SBAG, 72 pp., posted July 2012, by the Mars Exploration Program Analysis Group (MEPAG) at http://mepag.jpl.nasa.gov/reports/.

[3] Hollingsworth, J.L., Haberle, R.M, Barnes, J.R., Bridger, A.F.C., Pollack, J.B, Lee, H. and Schaeffer, J. (1996). Orographic control of storm zones on Mars. *Nature*, **380,** 413–416. doi: 10.1038/380413a0

[4] Kahre, M.A., Hollingsworth, J.L., Haberle, R.M., Wilson, R.J. (2015). Couling the Mars dust and water cycles: The importance of radiative-dynamic feedbacks during northern hemisphere summer. *Icarus*, **260,** 477-480. doi: 10.1016/j.icarus.2014.07.017

[5] Kahre, M.A., Haberle, R.M., Hollingsworth, J.L. Wolff, M.J. (2020). MARCI-observed clouds in the Hellas Basin during northern hemisphere summer on Mars: Interpretation with the NASA/Ames legacy Mars global climate model. *Icarus*, **338**, 113512. doi: 10.1016/j.icarus.2019.113512

[6] Kass, D. M., Schofield, J. T., Michaels, T. I., Rafkin, S. C. R., Richardson, M. I., and Toigo, A. D. ( 2003), Analysis of atmospheric mesoscale models for entry, descent, and landing, *J. Geophys. Res.*, 108, 8090, doi:10.1029/2003JE002065, E12.

[7] Newman, C.E., J. Gómez-Elvira, M. Marin, S. Navarro, J. Torres, M.I. Richardson, J.M. Battalio, S.D. Guzewich, R. Sullivan, M. de la Torre, A.R. Vasavada, and N.T. Bridges (2017), Winds measured by the Rover Environmental Monitoring System (REMS) during the Mars Science Laboratory (MSL) Bagnold Dunes Campaign and comparison with numerical modeling using MarsWRF, Icarus, 291, 203-231, https://doi.org/10.1016/j.icarus.2016.12.016.

[8] Rafkin, S. C. R., 2009: A positive radiative-dynamic feedback mechanism for the maintenance and growth of Martian dust storms, J. Geophys. Res., 114, E01009, doi:10.1029/2008JE003217.

[9] Spiga, A., Faure, J., Madeleine, J.-B., Määttänen, A., and Forget, F. ( 2013). Rocket dust storms and detached dust layers in the Martian atmosphere, *J. Geophys. Res. Planets*, 118, 746–767, doi:10.1002/jgre.20046

[10] Gillespie, H. E., Greybush, S. J., & Wilson, R. J. ( 2020). An investigation of the encirclement of Mars by dust in the 2018 global dust storm using EMARS. *Journal of Geophysical Research: Planets*, 125, e2019JE006106. https://doi.org/10.1029/2019JE006106

[11] Bertrand, T., Wilson, R. J., Kahre, M. A., Urata, R., & Kling, A. ( 2020). Simulation of the 2018 global dust storm on Mars using the NASA Ames Mars GCM: A multitracer approach. *Journal of Geophysical Research: Planets*, 125, e2019JE006122. https://doi.org/10.1029/2019JE006122

[12] Newman, C.E. and co-authors (2020), Toward a more realistic simulation and prediction of dust storms on Mars, a White Paper submitted to the 2023-2032 Planetary Science and Astrobiology Decadal Survey.

[13] Grotzinger, J. P., & Milliken, R. E. (2012). The sedimentary rock record of Mars: Distribution, origins, and global stratigraphy. In J. P. Grotzinger & R. E. Milliken (Eds.), Sedimentary Geology of Mars (SEPM Speci, pp. 1–48). Retrieved from http://sp.sepmonline.org/content/sepsp102/1.toc

[14] Greeley, R., R. N. Leach, S. H. Williams, B. R. White, J. B. Pollack, D. H. Krinsley, and J. R. Marshall (1982), Rate of wind abrasion on Mars, J. Geophys. Res., 87(B12), 10,009-10,024.

[15] Armstrong, J., and C. Leovy (2005), Long term wind erosion on Mars, Icarus, 176(1), 57–74, doi:10.1016/j.icarus.2005.01.005.

[16] Golombek, M. P., N. H. Warner, V. Ganti, M. P. Lamb, T. J. Parker, R. L. Fergason, and R. Sullivan (2014), Small crater modification on Meridiani Planum and implications for erosion rates and climate change on Mars, J. Geophys. Res. E Planets, 119(12), 2522–2547, doi:10.1002/2014JE004658.

[17] Farley, K.A., Malespin, C., Mahaffy, P., Grotzinger, J.P., Vasconcelos, P.M., Milliken, R.E., Malin, M., Edgett, K.S., Pavlov, A.A., Hurowitz, J.A. and Grant, J.A., 2014. In situ radiometric and exposure age dating of the Martian surface. science, 343(6169), p.1247166.

[18] Day, M., & Dorn, T. ( 2019). Wind in Jezero crater, Mars. *Geophysical Research Letters*, 46, 3099– 3107. https://doi.org/10.1029/2019GL082218

[19] Bridges, N. T., Bourke, M. C., Geissler, P. E., Banks, M. E., Colon, C., Diniega, S., ... & Mellon, M. T. (2012). Planet-wide sand motion on Mars. Geology, 40(1), 31-34.

[20] Chojnacki, M., Banks, M. E., Fenton, L. K., & Urso, A. C. (2019). Boundary condition controls on the high-sand-flux regions of Mars. Geology, 47(5), 427-430.

[21] Baker, M. M., Newman, C. E., Lapotre, M. G. A., Sullivan, R., Bridges, N. T., & Lewis, K. W. (2018). Coarse sediment transport in the modern Martian environment. Journal of Geophysical Research: Planets, 123(6), 1380-1394.

[22] Greeley, R., Leach, R., White, B., Iversen, J., & Pollack, J. (1980). Threshold windspeeds for sand on Mars: Wind tunnel simulations. Geophysical Research Letters, 7(2), 121-124.

[23] Shao, Y., & Lu, H. (2000). A simple expression for wind erosion threshold friction velocity. Journal of Geophysical Research: Atmospheres, 105(D17), 22437-22443.

[24] Kok, J. F. (2010). Difference in the wind speeds required for initiation versus continuation of sand transport on Mars: Implications for dunes and dust storms. Physical Review Letters, 104(7), 074502.

[25] MEPAG NEX-SAG Report (2015), Report from the Next Orbiter Science Analysis Group (NEX-SAG), *Chaired by* B. Campbell and R. Zurek, 77 pages posted December, 2015 by the Mars Exploration Program Analysis Group (MEPAG) at http://mepag.nasa.gov/reports.cfm.

[26] MEPAG ICE-SAG Final Report (*2019*), Report from the Ice and Climate Evolution Science Analysis group (ICE-SAG), *Chaired by* S. Diniega and N. E. Putzig, 157 pages posted 08 July 2019, by the Mars Exploration Program Analysis Group (MEPAG) at http://mepag.nasa.gov/reports.cfm.

[27] National Research Council. 2011. *Vision and Voyages for Planetary Science in the Decade 2013-2022*. Washington, DC: The National Academies Press. https://doi.org/10.17226/13117.

[28] Martínez, G. M., et al. "The modern near-surface Martian climate: A review of in-situ meteorological data from Viking to Curiosity." *Space Science Reviews* 212.1-2 (2017): 295-338.

[29] Spiga, Aymeric, et al. "Atmospheric science with InSight." *Space Science Reviews* 214.7 (2018): 109.

[30] Manfredi, J. A., et al. "MEDA: The Mars Environmental Dynamics Analyzer. A suite of sensors for the Mars 2020 mission". *Space Science Reviews* (2020; under review).

[31] Banfield, D., Spiga, A., Newman, C. *et al.* The atmosphere of Mars as observed by InSight. *Nat. Geosci.* 13, 190–198 (2020). https://doi.org/10.1038/s41561-020-0534-0

[32] Benna M, Bougher SW, Lee Y, et al. Global circulation of Mars' upper atmosphere. *Science*. 2019;366(6471):1363-1366. doi:10.1126/science.aax1553

[33] Lellouch, E., Rosenqvist, J., Goldstein, J.J., Bougher, S.W., Paubert, G., 1991. First absolute wind measurements in the middle atmosphere of Mars. Astrophys. J. 383, 401–406.

[34] Cavalié, T., F. Billebaud, T. Encrenaz, M. Dobrijevic, J. Brillet, F. Forget and E. Lellouch (2008), Vertical temperature profile and mesospheric winds retrieval on Mars from CO millimeter observations - Comparison with general circulation model predictions. A&A, 489 2, 795-809, DOI: https://doi.org/10.1051/0004-6361:200809815.

[35] Moreno, R., E. Lellouch, F. Forget, T. Encrenaz, S. Guilloteau, E. Millour (2009), Wind measurements in Mars' middle atmosphere: IRAM Plateau de Bure interferometric CO observations, Icarus 201, 2, pp. 549-563. https://doi.org/10.1016/j.icarus.2009.01.027.

[36] Sonnabend, G., Sornig, M., Krötz, P. J., Schieder, R. T., and Fast, K. E. (2006), High spatial resolution mapping of Mars mesospheric zonal winds by infrared heterodyne spectroscopy of CO2, *Geophys. Res. Lett.*, 33, L18201, doi:10.1029/2006GL026900.

[37] Wang, H., and Ingersoll, A. P. ( 2003), Cloud-tracked winds for the first Mars Global Surveyor mapping year, *J. Geophys. Res.*, 108, 5110, doi:10.1029/2003JE002107, E9.

[38] McConnochie, T. H., J.F. Bell III, D. Savransky, M.J. Wolff, A.D. Toigo, H. Wang, M.I. 818 Richardson, and P.R. Christensen (2010), THEMIS-VIS Observations of Clouds in the Martian Mesosphere: Altitudes, Wind Speeds, and Decameter-Scale Morphology, *Icarus*, 820 *210*, 545-565, doi 10.1016/j.icarus.2010.07.021

[39] Hernández-Bernal, J., Sánchez-Lavega, A., del Río-Gaztelurrutia, T., Hueso, R., Cardesín-Moinelo, A., Ravanis, E. M., et al. ( 2019). The 2018 Martian global dust storm over the South Polar Region studied with MEx/VMC. *Geophysical Research Letters*, 46, 10330–10337. https://doi.org/10.1029/2019GL084266

[40] Levrard, B., Forget, F., Montmessin, F. et al. (2004). Recent ice-rich deposits formed at high latitudes on Mars by sublimation of unstable equatorial ice during low obliquity. Nature 431, 1072–1075,. https://doi.org/10.1038/nature03055

[41] Newman, Claire E.; Lewis, Stephen R. and Read, Peter L. (2005). The atmospheric circulation and dust activity in different orbital epochs on Mars. Icarus, 174(1) pp. 135–160.

[42] Mischna, M. A., V. Baker, R. Milliken, M. Richardson, and C. Lee (2013), Effects of obliquity and water vapor/trace gas greenhouses in the early martian climate, J. Geophys. Res. Planets, 118, 560–576, doi:10.1002/jgre.20054.

[43] Lewis, S. R., Read, P. L., Conrath, B. J., Pearl, J. C., & Smith, M. D. (2007). Assimilation of thermal emission spectrometer atmospheric data during the Mars Global Surveyor aerobraking period. *Icarus*, 192, 327–347. https://doi.org/10.1016/j.icarus.2007.08.009

[44] Hoffman, M. J., Greybush, S. J., John Wilson, R., Gyarmati, G., Hoffman, R. N., Kalnay, E.,…Szunyogh, I. (2010). An ensemble Kalman filter data assimilation system for the Martian atmosphere: Implementation and simulation experiments. *Icarus*, 209, 470–481. https://doi.org/10.1016/j.icarus.2010.03.034

[45] Lee, C., Lawson, W. G., Richardson, M. I., Heavens, N. G., Kleinböhl, A., Banfield, D.,…Toigo, A. D. (2009). Thermal tides in the Martian middle atmosphere as seen by the Mars Climate Sounder. *Journal of Geophysical Research*, 114, E03005. https://doi.org/10.1029/2008JE003285

[46] Greybush, S. J., Wilson, R. J., Hoffman, R. N., Hoffman, M. J., Miyoshi, T., Ide, K.,…Kalnay, E. (2012). Ensemble Kalman filter data assimilation of Thermal Emission Spectrometer temperature retrievals into a Mars GCM. *Journal of Geophysical Research*, 117, E11008. https://doi.org/10.1029/2012JE004097

[47] Greybush, S.J., H.E. Gillespie, R. J. Wilson (2019), Transient eddies in the TES/MCS Ensemble Mars Atmosphere Reanalysis System (EMARS), Icarus, 10.1016/j.icarus.2018.07.001, 317, 158-181.

[48] Navarro, T., Forget, F., Millour, E., and Greybush, S. J. (2014). Detection of detached dust layers in the Martian atmosphere from their thermal signature using assimilation. Geophys. Res. Lett., 41, 6620-6626.

[49] Smith, M. D., Pearl, J. C., Conrath, B. J., & Christensen, P. R. (2001). Thermal emission spectrometer results: Mars atmospheric thermal structure and aerosol distribution. *Journal of Geophysical Research*, 106, 23,929– 23,945. https://doi.org/10.1029/2000JE001321

[50] Kleinböhl, A., et al. ( 2009), Mars Climate Sounder limb profile retrieval of atmospheric temperature, pressure, and dust and water ice opacity, *J. Geophys. Res.*, 114, E10006, doi:10.1029/2009JE003358.

[51] Kleinböhl, A., A. J. Friedson, and J. T. Schofield, Two-dimensional radiative transfer for the retrieval of limb emission measurements in the Martian atmosphere, *J. Quant. Spectrosc. Radiat. Transfer, 187*, 511-522, doi:10.1016/j.jqsrt.2016.07.009, 2017.

[52] Prince JL, Desai PN, Queen EM, Grover MR. Mars phoenix entry, descent, and landing simulation design and modeling analysis. Journal of Spacecraft and Rockets. 2011 Sep;48(5):756-64.

[53] Dust in the Atmosphere of Mars and its Impact on Human Exploration, edited by J.S. Levine, D. Winterhalter, and R.L. Kerschmann, 2018, Cambridge Scholars Publishing.

[54] McClean, J.B. and W.T. Pike (2017), Estimation of the Saltated Particle Flux at the Mars 2020 In-Situ Resource Utilization Experiment (MOXIE) Inlet, presented at the Dust in the Atmosphere of Mars and Its Impact on Human Exploration Workshop, Abstract #6025.

[55] Heavens, N.G., Kleinböhl, A., Chaffin, M.S. *et al.* Hydrogen escape from Mars enhanced by deep convection in dust storms. *Nat Astron* 2, 126–132 (2018). https://doi.org/10.1038/s41550-017-0353-4

[56] Chaffin, M., Deighan, J., Schneider, N. *et al.* Elevated atmospheric escape of atomic hydrogen from Mars induced by high-altitude water. *Nature Geosci* 10, 174–178 (2017). https://doi.org/10.1038/ngeo2887

[57] Fedorova, A.A. et al. (2020), Stormy water on Mars: The distribution and saturation of atmospheric water during the dusty season, Science, DOI: 10.1126/science.aay9522.

[58] Montabone, L. and N.G. Heavens et al. (2020), Observing Mars from Areostationary Orbit: Benefits and Applications, a White Paper submitted to the 2023-2032 Planetary Science and Astrobiology Decadal Survey.

[59] Cremons, D.R., Abshire, J.B., Sun, X. *et al.* Design of a direct-detection wind and aerosol lidar for mars orbit. *CEAS Space J* **12,** 149–162 (2020). https://doi.org/10.1007/s12567-020-00301-z

[60] Waters, J.W., Froidevaux, L., Harwood, R.S., Jarnot, R.F., Pickett, H.M., Read, W.G., Siegel, P.H., Cofield, R.E., Filipiak, M.J., Flower, D.A., Holden, J.R., Lau, G.K., Livesey, N.J.,

Manney, G.L., Pumphrey, H.C., Santee, M.L., Wu, D.L., Cuddy, D.T., Lay, R.R., Loo, M.S., Perun, V.S., Schwartz, M.J., Stek, P.C., Thurstans, R.P., Chandra, K.M., Chavez, M.C., Chen, G., Boyles, M.A., Chudasama, B.V., Dodge, R., Fuller, R.A., Girard, M.A., Jiang, J.H., Jiang, Y., Knosp, B.W., LaBelle, R.C., Lam, J.C., Lee, K.A., Miller, D., Oswald, J.E., Patel, N.C., Pukala, D.M., Quintero, O., Scaff, D.M., Snyder, W.V., Tope, M.C., Wagner, P.A., Walch, M.J., 2006. The earth observing system microwave limb sounder (EOS MLS) on the Aura satellite. IEEE Trans. Geosci. Remote Sens. 44 (5), 1075–1092. https://doi.org/10.1109/TGRS.2006.873771.

[61] Read, W. G.,, L. K. Tamppari, N. J. Livesey, R. T. Clancy, F. Forget, P. Hartogh, S. C. R. Rafkin, G. Chattopadhyay, 2018. Retrieval of wind, temperature, water vapor and other trace constituents in the Martian atmosphere, *Plan. and Sp. Sci.*, doi.org/10.1016/j.pss.2018.05.004..

[62] Gordley, L.L. and B.T. Marshall, "Doppler wind and temperature sounder: new approach using gas filter radiometry," J. Appl. Rem. Sens. 5(1) 053570 (1 January 2011) https://doi.org/10.1117/1.3666048